\documentclass[twocolumn,showpacs,preprintnumbers, 
               eqsecnum,nofootinbib]{revtex4}
\usepackage{graphicx, fancybox}
\usepackage{amsmath,amssymb}
\usepackage[dvipdfm, colorlinks=true, pdfstartview=FitV, linkcolor=red, citecolor=blue, urlcolor=blue]{hyperref}
\usepackage{color}

\newcommand{\vect}[1]{\mathbf{#1}}

\newcommand{\sgn}{\mathrm{sgn}}
\newcommand{\Slash}[1]{\ooalign{\hfil/\hfil\crcr$#1$}}

\newcommand{\vp}{\vect{p}}
\newcommand{\vk}{\vect{k}}
\newcommand{\vx}{\vect{x}}
\newcommand{\vy}{\vect{y}}

\newcommand{\vgamma}{{\boldsymbol \gamma}}
\newcommand{\cp}{g}

\newcommand{\comment}[1]{}

\newcommand{\jind}{j_{\mathrm {ind}}}
\newcommand{\etaind}{\eta_{\mathrm {ind}}}

\newcommand{\Tc}{{\mathrm T}_C}
\newcommand{\R}{R}
\newcommand{\A}{A}
\newcommand{\nf}{n_F}
\newcommand{\nb}{n_B}
\newcommand{\mf}{m_q}
\newcommand{\mb}{m_g}
\newcommand{\Cf}{C_f}
\newcommand{\Kzero}{K^{\text{(0)}}}
\newcommand{\Kone}{K^{\text{(1)}}}
\newcommand{\zetaf}{\zeta_q}
\newcommand{\zetab}{\zeta_g}
\newcommand{\VC}{ \varXi }
\newcommand{\V}{\tilde{\varGamma}}

\newcommand{\thetac}{\theta_C}

\begin{document}

\preprint{KUNS-2440} 

\title{Ultrasoft fermion mode and off-diagonal Boltzmann equation in quark-gluon plasma at high temperature}

\author{Daisuke Satow}
\email{d-sato@ruby.scphys.kyoto-u.ac.jp}
\affiliation{Department of Physics, Faculty of Science, Kyoto University, Kitashirakawa Oiwakecho, Sakyo-ku, Kyoto 606-8502, Japan}
\affiliation{Institut de Physique Th\'eorique, Orme des Merisiers batiment 774 Point courrier 136 CEA/DSM/IPhT, CEA/Saclay, F-91191 Gif-sur-Yvette Cedex, France}

\begin{abstract} 
We derive the generalized Boltzmann equation (GBE) near equilibrium from the Kadanoff-Baym equation for quark excitation with ultrasoft momentum ($\sim\cp^2T$, $\cp$: coupling constant, $T$: temperature) in quantum chromodynamics (QCD) at extremely high $T$, and show that the equation is equivalent to the self-consistent equation derived in the resummed perturbation scheme used to analyze the quark propagator.
We obtain the expressions of the dispersion relation, the damping rate, and the strength of a quark excitation with ultrasoft momentum by solving the GBE.
We also show that the GBE enables us to obtain the equation determining the $n$-point function containing a pair of quarks and $(n-2)$ gluon external lines whose momenta are ultrasoft.
\end{abstract}

\date{\today}

\maketitle

%%%%%%%%%%%%%%%%%%%%%%%%%%%%%%%%%%%%%%%%%%%%%%%%%%%%%%%%%%%%%%%%%%
\section{Introduction}
\label{sec:intro} 

Theories containing fermion and boson such as the Yukawa model, quantum electrodynamics (QED), and especially quantum chromodynamics (QCD) suggest that the plasmas described by these theories including QED plasma and the quark-gluon plasma~\cite{QGP} at so high temperature ($T$) that the particle masses are negligible are multi-scale systems.
For example, the average inter-particle distance is of order $ T^{-1}$, the Debye screening length, $(\cp T)^{-1}$ ($\cp$: coupling constant)~\cite{plasmon}, and the mean free path of the quark and the gluon, $(\cp^2T)^{-1}$~\cite{lebedev, quark-gluon-damping}, respectively.
These momentum scales, $T$, $\cp T$, and $\cp^2T$, are called hard, soft, ultrasoft, respectively.
Due to the existence of these multiple scales, analyzing the spectral properties of collective excitations at finite temperature in these fermion-boson systems is a quite nontrivial task even at weak coupling ($\cp\ll 1$) regime in general.
In the soft scale, the well-established perturbation theory known as the hard thermal loop (HTL) approximation~\cite{HTL} is applicable, and the analysis using that approximation suggests the existence of the plasmon~\cite{plasmon}, which is a bosonic excitation and well-known in nonrelativistic plasma.
In addition to the bosonic excitation, a fermionic excitation called plasmino~\cite{plasmino} is also suggested to exist, owing to the fact that the particle masses are negligible compared with $T$. 
By contrast, the HTL approximation can not be used in the ultrasoft scale due to an infrared singularity known as pinch singularity, and thus a novel resummed perturbation theory, which regularizes the singularity, needs to be developed to obtain the correct leading-order result~\cite{lebedev,ultrasoft-QED,susy-kinetic}.
Due to the same reason, the resummed perturbation theory is also necessary in the calculation of the bosonic propagator~\cite{blaizot-ultrasoft} and the transport coefficient~\cite{transport-perturbation,transport-kinetic,transport-2PI,transport-3PI}.
The analysis using the resummed perturbation revealed the existence of a novel quark excitation (ultrasoft fermion mode) whose momentum is ultrasoft~\cite{lebedev} in addition to the plasmino~\cite{plasmino}, and the expression of the pole position and the strength of that excitation was obtained by solving the self-consistent equation derived in the resummed perturbation, in the Yukawa model and QED~\cite{ultrasoft-QED}.
We note that results which suggest the existence of the ultrasoft fermion mode were also obtained by other analyses, in which the Schwinger-Dyson equation~\cite{S-Deq}, the Nambu-Jona-Lasino model~\cite{3peak-NJL}, and effective models in which the boson has finite bare mass~\cite{3peak} are used. 
On the other hand, in QCD, the self-consistent equation has not been solved, and thus the expressions of the pole position and the strength of the ultrasoft fermion mode have not been obtained.
In this paper, we extend the analysis in the Yukawa model and QED~\cite{ultrasoft-QED} to QCD, and obtain the expression of the pole position and the strength of that excitation.

Now we introduce another point of view: each perturbation scheme is equivalent to different kinetic equation near equilibrium.
In fact, the HTL analysis of the soft boson (fermion) propagator is equivalent to the usual (generalized) collisionless kinetic equation called (generalized) Vlasov equation~\cite{blaizot-HTL}.
Here the generalized Vlasov equation means the collisionless kinetic equation in which there is a fermionic background field in place of bosonic one, in contrast to the usual Vlasov equation case. 
On the other hand, it was shown that the resummation scheme used in the analysis of the gluon propagator whose momentum is ultrasoft, is equivalent to the Boltzmann equation~\cite{blaizot-ultrasoft}.
It reflects the fact that the interaction among the particles can not be neglected when the space-time scale is comparable with or much larger than the mean free path, which is of order $(\cp^2T)^{-1}$ in the case of the hard quark and gluon~\cite{lebedev,quark-gluon-damping}.
We note that this discussion explains the reason why the resummed perturbation, which takes into account the interaction effect among the hard particles, is necessary in the analysis of the ultrasoft momentum region. 
The resummaton scheme used in the analysis of the ultrasoft fermion propagator~\cite{lebedev,ultrasoft-QED} is also interpreted as a generalized version of the Boltzmann equation (GBE) near equilibrium in the case of the Yukawa model and QED~\cite{kinetic-offdiagonal}.
In this paper, we derive the GBE that is equivalent to the self-consistent equation derived in the resummed perturbation~\cite{lebedev} in QCD. 

The aim of this paper is twofold: One is to obtain the expression of the dispersion relation, the damping rate, and the strength of the ultrasoft fermion mode in QCD. 
The other is, to derive the GBE that is equivalent to the self-consistent equation obtained in the resummed perturbation~\cite{lebedev} from the Kadanoff-baym equation~\cite{kadanoff-baym, blaizot-review} in QCD. 
The derivation helps us to establish the foundation of the resummed perturbation scheme~\cite{lebedev}, and to obtain the kinetic interpretation of the procedure of that scheme.
As a by-product, the derivation using the Kadanoff-Baym equation enables us to evaluate also the nonlinear response that is caused by the average gluon field, not only the linear response caused by the quark average field~\cite{blaizot-review, blaizot-HTL, blaizot-ultrasoft}.
Due to this merit, we can analyze the $n$-point function whose external lines are $(n-2)$ ultrasoft gluons and a pair of ultrasoft quark, not only the ultrasoft quark propagator.
This paper is an generalization of Ref.~\cite{kinetic-offdiagonal, ultrasoft-QED}, from the Yukawa model and QED to QCD, and also an extension of Ref.~\cite{blaizot-ultrasoft}, from the gluon propagator to the quark propagator.
Since there is self-coupling of the gluon and the color structure of the off-diagonal propagator, which is an important quantity in our analysis and will be introduced later, is complicated in general in QCD, the extension to QCD is nontrivial and thus it deserves a paper.

This paper is organized as follows:
We derive the GBE near equilibrium that is equivalent to the self-consistent equation obtained in the resummed perturbation theory~\cite{lebedev, ultrasoft-QED} in linear response regime in Sec.~\ref{sec:kinetic}, in a similar method to the case of the Yukawa model and QED~\cite{kinetic-offdiagonal}.
We also show that the Ward-Takahashi identity is satisfied, and that the identity can be derived from the conservation of the color current.
In Sec.~\ref{sec:fermion-mode}, we analyze the properties of the ultrasoft fermion mode such as dispersion relation, damping rate, and residue in QCD, by solving the GBE in some momentum region as in the case of the Yukawa model and QED~\cite{ultrasoft-QED}.
Section~\ref{sec:higher} is devoted to obtaining the equation determining the $n$-point function whose external lines are a pair of quarks and $(n-2)$ gluons with ultrasoft momenta.
We summarize our paper and give concluding remarks in Sec.~\ref{sec:summary}.
In Appendix~\ref{app:longitudinal}, we show that the longitudinal component of the off-diagonal propagator introduced in Sec.~\ref{sec:kinetic} is much smaller than the transverse component of that quantity.
We give a detailed derivation of Eqs.~(\ref{eq:EOM-result}) and (\ref{eq:WT-calculation}) in Appendix~\ref{app:off-diagonal}.
Appendix~\ref{app:ultrasoft-mode} is devoted to derivation of Eq.~(\ref{eq:fermion-mode-result}).

%%%%%%%%%%%%%%%%%%%%%%%%%%%%%%%%

\begin{figure*}[t]
\begin{center}
\includegraphics[width=0.60\textwidth]{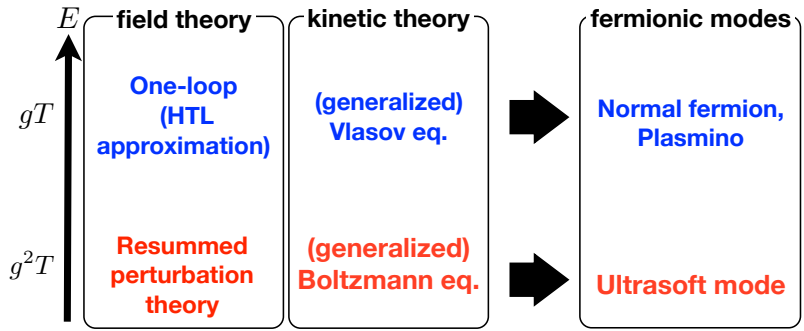}
\caption{The correspondence between the diagrammatic methods and the kinetic equations, and the fermionic modes that result from the analysis using the two methods.
The vertical axis denote the energy scale.}
\label{fig:intro}
\end{center}
\end{figure*}

\begin{figure}[t]
\begin{center}
\includegraphics[width=0.35\textwidth]{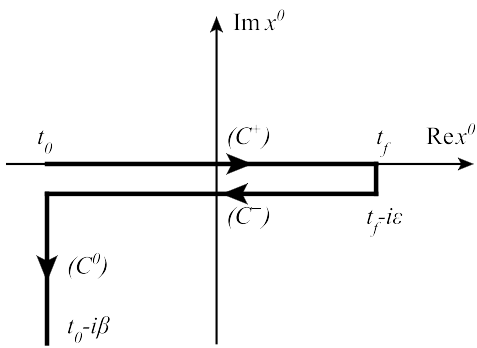}
\caption{The contour $C$ in the complex $x^0$ plane.}
\label{fig:contour}
\end{center}
\end{figure}

%%%%%%%%%%%%%%%%%%%%%%%%%%%%%%%%%%%%%%%%%%%%%%%%%%%%%%%%%%%%%%%%%%
\section{Generalized Boltzmann equation}
\label{sec:kinetic}

In this section, we derive the GBE that is equivalent to the self-consistent equation obtained in the resummed perturbation theory~\cite{lebedev}, from the Kadanoff-Baym equation~\cite{kadanoff-baym,blaizot-review} by using the background field gauge method~\cite{background-field-gauge}, which enables us to obtain the resultant equations in a gauge-covariant way.
We also check that the resultant kinetic equation satisfies the Ward-Takahashi identity, which is a necessary condition of the $SU(N)$ gauge invariance.

%%%%%%%%%%%%%%%%%%%%%%%%%%%%%%%%
\subsection{Background field gauge method}
\label{sec:background-gauge}

Let us introduce the background field gauge method~\cite{background-field-gauge,blaizot-review}.
In this method, the following generating functional is used:
\begin{align} 
\label{eq:generating-functional}
\begin{split}
&\tilde{Z}[j, \eta,\overline{\eta}; A, \varPsi,\overline{\varPsi}]=\int [{\cal D}{a}][{\cal D}\overline{\psi}][{\cal D}\psi]
[{\cal D}\overline{\zeta}][{\cal D}\zeta]  \\
&\times\exp\Bigl(i\int_C d^4x({\cal L}[A^a_\mu+a^a_\mu, \varPsi+\psi, \overline{\varPsi}+\overline{\psi}]+{\cal L}_{\text {FP}}\\
&-(j^a_\mu a^{a \mu}+\overline{\psi}\eta+\overline{\eta}\psi)) \Bigr), 
\end{split}
\end{align}
 where the time integral is defined using the closed-time-path formalism~\cite{blaizot-review,lebellac}: $\int_C d^4x\equiv \int_C dx^0\int d^3 \vx$, where $\int_C dx^0$ is the integral along the contour $C$, which is defined in Fig.~\ref{fig:contour} and consists of $C^+$, $C^-$, and $C^0$.  
Here $t_f$ is larger than any time we consider.
$A^a_\mu$ and $a^a_\mu$ are the vector fields, $\varPsi$ ($\overline{\varPsi}$) and $\psi$ ($\overline{\psi}$) the (anti-) spinor fields, $\zeta$ ($\overline{\zeta}$) the (anti-) ghost field, $j^a_\mu$ the external current, and $\eta$ ($\overline{\eta}$) the external (anti-) fermion source, respectively.

The Lagrangian of QCD is defined as
\begin{align}
&{\cal L}[a,\psi,\overline{\psi}]= i\sum^{N_f}_{j=1}\overline{\psi}_j\Slash{D}[a]\psi_j-\frac{1}{4}F^{\mu\nu a}[a]F^a_{\mu\nu}[a], 
\end{align} 
where $F^a_{\mu\nu}\equiv \partial_\mu a^a_\nu-\partial_\nu a^a_\mu-\cp f^{abc}a^b_\mu a^c_\nu$ is the field strength, $D_\mu[a]\equiv \partial_\mu +i\cp a^a_\mu t^a$ the covariant derivative in the fundamental representation, $t^a$ ($a=1,... N^2-1$) the generator of $SU(N)$ group in the fundamental representation, $f^{abc}$ the structure constant of $SU(N)$ group, respectively.
We note that $N_f$, the flavor number, and $N$, the color number, are not specified in this paper, because we want to see how the expressions of the properties of the ultrasoft mode, which will be analyzed in Sec.~\ref{sec:fermion-mode}, depend on $N$ and $N_f$.
In the real world, $N=3$.
Since we are considering the case that the quark current mass is negligible in every flavor, we drop the index for the flavor from now on.

We adopt the temporal gauge fixing, which was used in the analysis based on the resummed perturbation~\cite{lebedev}.
In this gauge-fixing, the Faddeev-Popov term is
\begin{align}
&{\cal L}_{\text {FP}}=-\lambda\frac{(\tilde{G}^a[a])^2}{2} 
- \overline{\zeta}^a\left(\frac{\partial_0}{\cp}\delta^{ab}-f^{adb}(A+a)^d_0\right)\zeta^b, 
\end{align}
where the gauge-fixing function is $\tilde{G}^a[a]= a^a_0$ with $\lambda\rightarrow \infty$, which is equivalent to the following constraint:
\begin{align}
\label{eq:temporal-gauge}
a^a_0=0. 
\end{align}
We note that the possible gauge-fixing dependence needs to be checked apart from the covariance with respect to the background gauge transformation, which will be introduced later.

In the background field gauge method, we impose the following conditions:
\begin{align}
\label{eq:impose}
\langle a^a_\mu\rangle=\langle\psi\rangle=\langle\overline{\psi}\rangle=0. 
\end{align}
$A^a_\mu$, $\varPsi$ ($\overline{\varPsi}$) become equal to the gluon and the (anti-) quark average field after imposing these conditions~\cite{background-field-gauge}.

We note that the action in Eq.~(\ref{eq:generating-functional}) is invariant under the following transformation~\cite{blaizot-review}: 
\begin{align} 
\label{eq:gauge-transform}
\begin{split}
\varPsi(x)\rightarrow h(x)\varPsi(x),~~~
\overline{\varPsi}(x)\rightarrow \overline{\varPsi}(x) h^\dagger(x),\\
\psi(x)\rightarrow h(x)\psi(x),~~~
\overline{\psi}(x)\rightarrow \overline{\psi}(x) h^\dagger(x),\\
A^a_\mu(x)t^a\rightarrow h(x)A^a_\mu(x)t^ah^\dagger(x) -\frac{i}{\cp}h\partial_\mu h^\dagger(x),\\
a^a_\mu(x)t^a\rightarrow h(x) a^a_\mu(x)t^a h^\dagger(x), \\
j^a_\mu(x)t^a \rightarrow h(x) j^a_\mu(x)t^a h^\dagger(x),\\
\eta(x)\rightarrow h(x)\eta(x),~~~
\overline{\eta}(x)\rightarrow \overline{\eta}(x) h^\dagger(x),\\
\zeta^a(x)t^a \rightarrow h(x) \zeta^a(x)t^a h^\dagger(x),\\
\overline{\zeta}^a(x)t^a \rightarrow h^\dagger(x)\overline{\zeta}^a(x)t^a h(x), 
\end{split}
\end{align} 
where $h(x)\equiv \exp[i\theta^a(x)t^a]$.

%%%%%%%%%%%%%%%%%%%%%%%%%%%%%%%%
\subsection{Derivation}
\label{ssc:kinetic-derivation}

We consider the following situation:
The system is at equilibrium whose temperature is $T$ before the initial time $t_0$.
Then, external quark, anti-quark, and gluon external sources disturb the system, and hence the system becomes non-equilibrium state.
In this paper, we focus on the case where the external fermion source is so weak that we need to retain only the contributions that are in the linear order in the magnitude of the fermionic average field $\varPsi$ to the induced fermion source, which will be introduced later.
Due to the linear response theory, analysis in such situation is equivalent to the computation of the fermion propagator with ultrasoft momentum at equilibrium.
We note however, that the induced fermion source contains the all order contributions in $A^a_\mu$ in our approximation, in which we have a good machinery to compute the $n$-point function whose external lines are two quarks and $(n-2)$ gluons and whose all momenta are ultrasoft, not only the ultrasoft fermion propagator. 
In this sense, we are going to analyze the region which is beyond the linear response regime. 
The derivation will be performed in a similar way to that in Ref.~\cite{blaizot-ultrasoft,kinetic-offdiagonal}:
We will apply the gradient expansion and the weak coupling approximation to the Kadanoff-Baym equation~\cite{kadanoff-baym,blaizot-review}.
We note that both of them are justified by the smallness of the coupling constant, as we will see later.

By performing a infinitesimal variation of the integral variable in Eq.~(\ref{eq:generating-functional}), we get the following equations:
\begin{align}
\label{eq:EOM-fermion}
\langle i\Slash{D}_x[A+a](\varPsi+\psi)(x)\rangle&=\eta(x).\\
\nonumber
 \langle (\tilde{D}_\nu[A+a])^{ab} F^{b \nu}_{~~\mu}[A+a]  &
 -\cp\langle(\overline{\varPsi}+\overline{\psi})t^a\gamma_\mu(\varPsi+\psi) \rangle \\
 \label{eq:EOM-boson}
-f^{abc}g_{0\mu}\langle\overline{\zeta}^b\zeta^c\rangle
 &=j^a_\mu(x). 
\end{align}
Here $\tilde{D}_\mu[a]\equiv \partial_\mu +i\cp a^a_\mu T^a$ is the covariant derivative in the adjoint representation, where $(T^a)_{bc}\equiv -if^{abc}$ is the generator of $SU(N)$ group in the adjoint representation.
By imposing Eq.~(\ref{eq:impose}), we get the following equations of motion for the average fields:
\begin{align}
\label{eq:meanfield-fermion}
i\Slash{D}_x[A]\varPsi(x)&=\eta(x)+\etaind(x),\\
\label{eq:meanfield-boson}
\tilde{D}^{ab}_{\nu x}[A] F^{b\nu\mu}[A](x)-\cp\overline{\varPsi}(x)t^a\gamma^\mu\varPsi(x)&= 
j^{a\mu}(x)+\jind^{a\mu}(x).
\end{align}
Here $\etaind$ and $\jind^{a\mu}$ are the induced fermion source and the induced color current defined as
\begin{align}
\label{eq:induced-current-fermion}
\etaind(x)&\equiv \cp t^a\langle \Slash{a}^a(x)\psi(x) \rangle,\\ 
\nonumber
\jind^{a\mu}(x)&\equiv \cp\langle \overline{\psi}t^a\gamma^\mu \psi\rangle
 + f^{abc}g_{0\mu}\langle\overline{\zeta}^b\zeta^c\rangle\\
 \nonumber
 &~~~+ \cp f^{abc}\varGamma^{\mu\rho\lambda\nu}\langle a^b_\nu\partial_\lambda a^c_\rho \rangle
 +\cp^2\hat\varGamma^{\mu\nu\rho\lambda}_{abcd}A^b_\nu\langle a^c_\rho a^d_\lambda\rangle \\
 &~~~+ \cp^2f^{abc}f^{cde}\langle a^b_\nu a^{d\mu}a^{e\nu}\rangle, 
\end{align}
where we have introduced $\varGamma^{\mu\rho\lambda\nu}\equiv 2g^{\mu\rho}g^{\lambda\nu}-g^{\mu\lambda}g^{\rho\nu}-g^{\mu\nu}g^{\rho\lambda}$ and 
\begin{align}
\begin{split}
\hat\varGamma^{\mu\nu\rho\lambda}_{abcd}&\equiv 
\frac{1}{2}[f^{eab}f^{ecd}\left(g^{\mu\rho}g^{\nu\lambda}-g^{\mu\lambda}g^{\nu\rho}\right)\\
&~~~+f^{eac}f^{edb}(g^{\mu\lambda}g^{\rho\nu}-g^{\mu\nu}g^{\lambda\rho})\\
&~~~+f^{ead}f^{ebc}(g^{\mu\nu}g^{\rho\lambda}-g^{\mu\rho}g^{\lambda\nu})].
\end{split}
\end{align} 
It was shown~\cite{blaizot-review} that Eqs.~(\ref{eq:meanfield-fermion}) and (\ref{eq:meanfield-boson}) transform under the gauge transformation given in Eq.~(\ref{eq:gauge-transform}) in a covariant way. 

By differentiating Eq.~(\ref{eq:EOM-fermion}) with respect to $j^{a\mu}(y)$ and Eq.~(\ref{eq:EOM-boson}) with respect to $\overline{\eta}(y)$, we get the following equations:
\begin{align}
\label{eq:EOM-propagator-fermion}
&i\Slash{D}_x[A]K^a_\mu(x,y)-\cp \gamma^\nu D^{ba}_{\nu\mu}(x,y)t^b\varPsi(x) 
=i\frac{\delta\etaind(x)}{\delta j^{\mu a}(y)},\\
\nonumber
&((\tilde{D}^2[A])^{ab}g^{\mu\nu} -(\tilde{D}^\mu[A] \tilde{D}^\nu[A])^{ab}+2i\cp F^{c\mu\nu}[A]T^c_{ab})_x\\
\label{eq:EOM-propagator-boson}
&\times K^{b}_{\nu}(y,x) \\
\nonumber
&-\cp\left(\overline{\varPsi}(x)t^a \gamma^\mu\langle \psi(x)\psi(y)\rangle 
+S(y,x)t^a\gamma^\mu\varPsi(x)\right)\\
\nonumber
&= i\frac{\delta\jind^{a\mu}(x)}{\delta \overline{\eta}(y)}, 
\end{align} 
where we have introduced the following propagators:
\begin{align} 
D^{ab}_{\mu\nu}(x,y)&\equiv \langle\Tc a^a_\mu(x) a^b_\nu(y)\rangle,\\
S(x,y)&\equiv \langle \Tc \psi(x)\overline{\psi}(y)\rangle,\\
K^a_\mu(x,y)&\equiv \langle\Tc \psi(x) a^a_\mu(y)\rangle.
\end{align}
Here $\Tc$ means the path-ordered product whose path is $C$:
\begin{align}
\nonumber
D^{ab}_{\mu\nu}(x,y)&= \thetac(x^0, y^0)D^{> ab}_{\mu\nu}(x,y)\\
&~~~+\thetac(y^0, x^0)D^{< ab}_{\mu\nu}(x,y),\\
S(x,y)&=  \thetac(x^0, y^0)S^>(x,y)-\thetac(y^0, x^0)S^<(x,y),\\
K^a_\mu(x,y)&= \thetac(x^0, y^0)K^{> a}_\mu(x,y)+\thetac(y^0, x^0)K^{< a}_\mu(x,y).
\end{align}
where we have introduced the step-function along the contour $C$, $\thetac(x,y)$, and the following functions:
\begin{align}
D^{> ab}_{\mu\nu}(x,y)&\equiv \langle a^a_\mu(x) a^b_\nu(y)\rangle,\\
D^{< ab}_{\mu\nu}(x,y)&\equiv \langle a^b_\nu(y) a^a_\mu(x)\rangle,\\
S^>(x,y)&\equiv \langle\psi(x) \overline{\psi}(y)\rangle,\\
S^<(x,y)&\equiv \langle\overline{\psi}(y)\psi(x)\rangle,\\
K^{> a}_\mu(x,y)&\equiv  \langle\psi(x) a^a_\mu(y)\rangle,\\
K^{< a}_\mu(x,y)&\equiv \langle a^a_\mu(y)\psi(x)\rangle.
\end{align}
In fact, $K^{> a}_\mu(x,y)$ coincides with $K^{< a}_\mu(x,y)$ in our approximation as in the case of the Yukawa model and QED~\cite{kinetic-offdiagonal}, so we simply write these two functions as $K^{ a}_\mu(x,y)$ from now on.
We call $K^a_\mu$ ``{\it{off-diagonal propagator}}'' since that quantity is the propagator between the quark and the gluon~\cite{kinetic-offdiagonal}.
We note that $K^a_\mu$ vanishes at equilibrium.
The off-diagonal propagator is the most important propagator in our analysis, which can be seen from the discussion in Sec.~\ref{ssc:equivalence}.
The transformation form of these propagators under the gauge transformation is given in Ref.~\cite{blaizot-review}.

From Eq.~(\ref{eq:temporal-gauge}), we get
\begin{align} 
D^{ab}_{0\mu}(x,y)&=D^{ab}_{\mu 0}(x,y)=0,\\
K^a_0(x,y)&=0.
\end{align}
Because of these constraints, we do not have to deal with the temporal components of these propagators.

By setting $x^0\in C^+$ and $y^0\in C^-$, and interchanging $x$ and $y$ in Eq.~(\ref{eq:EOM-propagator-boson}), we get
\begin{align}
\label{eq:EOM-propagator-2-fermion}
&i\Slash{D}_x[A]K^a_i(x,y)-\cp \gamma^j D^{<ba}_{j i}(x,y)t^b\varPsi(x) 
=i\frac{\delta\etaind(x)}{\delta j^{i a}(y)},\\
\nonumber
&((\tilde{D}^2[A])^{ab}g^{ij} -(\tilde{D}^i[A] \tilde{D}^j[A])^{ab}+2i\cp F^{cij}[A]T^c_{ab})_y\\
\nonumber
&~~~\times K^{b}_{j}(x,y) \\
\label{eq:EOM-propagator-2-boson}
&+\cp S^<(x,y)t^a\gamma^i\varPsi(y)
= i\frac{\delta\jind^{ai}(y)}{\delta \overline{\eta}(x)},
\end{align}
where we have neglected $\overline{\varPsi}(y)\langle\psi(y)\psi(x) \rangle$ because $\langle\psi(y)\psi(x) \rangle$ contains more than one average quark field.
The induced source terms are evaluated as follows~\cite{kinetic-offdiagonal}:
\begin{align} 
\nonumber
&\frac{\delta\etaind(x)}{\delta j^{i a}(y)}\\
\nonumber
&=\int_C d^4z\left( \varSigma(x,z)K^{ a}_i(z,y)+\VC^{bj}_{}(x,z) D^{ ba }_{ji}(z,y)  \right)\\ 
\nonumber
&=-i\int^\infty_{-\infty} d^4z\Bigl( \varSigma^\R(x,z)K^{ a}_i(z,y)\\
\label{eq:retarded-selfenergy-fermion}
&~~~+\VC^{\R bj}_{}(x,z) D^{< ba }_{ji}(z,y)   \Bigr), \\ 
\nonumber
&\frac{\delta\jind^{a i}(y)}{\delta \overline{\eta}(x)}\\
\nonumber
&=\int_C d^4z\left(\varPi^{abij}_{}(z,y)K^{b}_{j}(x,z)+ \VC^{a i}_{}(z,y) S(x,z)  \right)\\
\nonumber
&=-i\int^\infty_{-\infty} d^4z ( \varPi^{\A abij}_{}(z,y)K^{b}_{j}(x,z)\\
\label{eq:retarded-selfenergy-boson}
&~~~- \VC^{\R a i}_{}(z,y) S^<(x,z)  ).
\end{align}
Here $\varSigma$ is the quark self-energy, $\varPi^{ab}_{\mu\nu}$ the gluon self-energy, and $\VC^a_\mu$ the off-diagonal self-energy~\cite{kinetic-offdiagonal}, respectively, and they are decomposed as follows:
\begin{align}
\varSigma (x,y)&= \thetac(x^0, y^0)\varSigma^>(x,y)-\thetac(y^0, x^0)\varSigma^<(x,y),\\
\nonumber
\varPi^{ab}_{\mu\nu} (x,y)&= \thetac(x^0, y^0)\varPi^{ab>}_{\mu\nu}(x,y)\\
&~~~+\thetac(y^0, x^0)\varPi^{ab<}_{\mu\nu}(x,y),\\
\VC^a_\mu (x,y)&= \thetac(x^0, y^0)\VC^{a>}_\mu(x,y) +\thetac(y^0, x^0)\VC^{a<}_\mu(x,y).
\end{align}
We have also introduced the retarded quark self-energy $\varSigma^\R$, the advanced gluon self-energy $\varPi^{\A ab}_{\mu\nu}$, and the retarded off-diagonal self-energy $\VC^{\R a}_{\mu}$, which are defined as
\begin{align}
\varSigma^\R(x,y)&\equiv i \theta(x^0, y^0)[\varSigma^>(x,y)+\varSigma^<(x,y)], \\
\varPi^{\A ab}_{\mu\nu}(x,y)&\equiv -i\theta(y^0, x^0)[\varPi^{ab>}_{\mu\nu}(x,y)-\varPi^{ab<}_{\mu\nu}(x,y)] ,\\
\VC^{\R a}_{\mu}(x,y)&\equiv  i\theta(x^0, y^0)[\VC^{a>}_\mu(x,y)-\VC^{a<}_\mu (x,y)].
\end{align}
From Eqs.~(\ref{eq:EOM-propagator-2-fermion})$\sim$(\ref{eq:retarded-selfenergy-boson}), we have
\begin{align}
\label{eq:EOM-fermion-coordinate}
&i\Slash{D}_x[A]K^a_i(x,y)-\cp \gamma^j D^{<ba}_{j i}(x,y)t^b\varPsi(x) \\
\nonumber
&=\int^\infty_{-\infty} d^4z\Bigl( \varSigma^\R(x,z)K^{ a}_i(z,y)
+\VC^{\R bj}_{}(x,z) D^{< ba }_{j i}(z,y)   \Bigr), \\
\nonumber
&((\tilde{D}^2[A])^{ab}g^{ij} -(\tilde{D}^i[A] \tilde{D}^j[A])^{ab}+2i\cp F^{cij}[A]T^c_{ab})_y\\
\nonumber
&~~~\times K^{b}_{j}(x,y) \\ 
&+\cp S^<(x,y)t^a\gamma^i\varPsi(y)
\nonumber
= \int^\infty_{-\infty} d^4z ( \varPi^{\A abij}_{}(z,y)K^{b}_{j}(x,z)\\
\label{eq:EOM-boson-coordinate}
&~~~- \VC^{\R a i}_{}(z,y) S^<(x,z)  ).
\end{align} 

Now we perform the Wigner transformation~\cite{blaizot-review}, which is defined as
\begin{align}
f(k,X)\equiv\int d^4s e^{ik\cdot s}f\left(X+\frac{s}{2}, X-\frac{s}{2}\right),
\end{align}
where $s\equiv x-y$, $X\equiv(x+y)/2$, and $f(x,y)$ is an arbitrary function.
After this manipulation, Eqs.~(\ref{eq:EOM-fermion-coordinate}) and (\ref{eq:EOM-boson-coordinate}) become
\begin{align}
\nonumber
&\left(\Slash{k}+\frac{i}{2}\Slash{\partial}_X-\cp\Slash{A}^b(X)t^b\right)K^a_i(k,X)\\
\nonumber
&-\cp \gamma^j D^{<ba}_{j i}(k,X)t^b\varPsi(X) \\
\label{eq:EOM-wigner-fermion}
&= \varSigma^\R(k,X)K^{ a}_i(k,X)
+\VC^{\R bj}_{}(k,X) D^{< ba }_{j i}(k,X) , \\
\nonumber
&\Bigl\{[(-k^2+ik\cdot\partial_X)\delta_{ab}-2\cp k\cdot A^c(X) (T^c)^{ab}]g^{kj}\\
\nonumber
& -\Bigl\{\left[-k^k k^j +\frac{i}{2}(\partial^k_X k^j+k^k\partial^j_X)\right]\delta^{ab}\\
\nonumber
&-\cp(T^c)^{ab}(k^kA^{cj}(X)+A^{ck}(X)k^j)\Bigr\}\Bigr\} K^{b}_{j}(k,X) \\
&+\cp S^<(k,X)t^a\gamma^k\varPsi(X)
\label{eq:EOM-wigner-boson}
=  \varPi^{\A abkj}_{}(k,X)K^{b}_{j}(k,X)\\
\nonumber
&~~~- \VC^{\R a k}_{}(k,X) S^<(k,X)  .
\end{align} 
Here we have used the gradient expansion~\cite{blaizot-HTL,blaizot-ultrasoft} by assuming $k\gg \partial_X$, as in Ref.~\cite{kinetic-offdiagonal}, and set $i\rightarrow k$ in the latter equation.
This assumption is justified~\cite{kinetic-offdiagonal} by using the fact that $k\sim T$ and $\partial_X\sim\cp^2T$, which is valid since we are focusing on the ultrasoft energy region.
By multiplying Eq.~(\ref{eq:EOM-wigner-fermion}) by $(\Slash{k}+i\Slash{\partial}_X/2-\cp\Slash{A}^b(X)t^b-\varSigma^\R(k,X))$, multiplying Eq.~(\ref{eq:EOM-wigner-boson}) by the projection operator into the transverse component $P^{T}_{ik}(k)$, which is defined as $P^T_{ \mu\nu}(k)\equiv g_{\mu i}g_{\nu j}(\delta_{ij}-\hat{k}_i\hat{k}_j)$ with $\hat{k}_i\equiv k_i/|\vk|$, and subtracting the latter from the former, we obtain 
\begin{align}
\label{eq:EOM-momentum} 
\begin{split}
&\left(2ik\cdot \partial_X-2\cp k\cdot A^b(X)t^b-\{\Slash{k},  \varSigma^\R(k,X)\} \right)K^a_i(k,X)\\
&-2\cp k\cdot A^c(T_c)^{ab}K^b_i(k,X)\\
&+P^T_{ik}(k)\varPi^{\A abkj}_{}(k,X)K^b_j(k,X) \\
&=\cp( \Slash{k}D^{< ba }_{j i}(k,X)+P^T_{ji}(k)\delta_{ab}S^<(k))\tilde{\varGamma}^{bj}(k,X),
\end{split}
\end{align} 
with $\cp\tilde{\varGamma}^a_\mu(k,X)\equiv \cp t^a\gamma_\mu \varPsi(X)+\VC^{\R a}_\mu(k,X)$. 
Here we have neglected the longitudinal component of $K^a_i(k,X)$, which is much smaller than the transverse one as is shown in Appendix~\ref{app:longitudinal}, and used $P^{T}_{ik}(k) K^{a k}(k,X)\simeq -K^{a}_{i}(k,X)$. 

The diagonal propagators in the right-hand side can be replaced by that in the free limit at equilibrium since we are considering the case where the system is near the equilibrium so that the induced fermion source contains only one average fermion field ($\varPsi$)~\cite{kinetic-offdiagonal}:
\begin{align}
\label{eq:freepropagator-fermion}
S^{0<}(k)&= \Slash{k}\rho^0(k)\nf(k^0),\\
\label{eq:freepropagator-boson}
D^{0< ab}_{ij}(k)&= \delta_{ab} \rho^0(k)\nb(k^0)P^T_{ij}(k) , 
\end{align}
where $\rho^0(k)\equiv  2\pi\sgn(k^0)\delta(k^2)$ is the free spectral function and $\nf(k^0)\equiv (e^{k^0/T}+1)^{-1}$ ($\nb(k^0)\equiv (e^{k^0/T}-1)^{-1}$) is the fermion (boson) distribution function at equilibrium.
We note that only $k$ satisfying $k^2=0$ contributes to the fermion induced source since $\delta(k^2)$ appears in the right-hand side of Eq.~(\ref{eq:EOM-momentum}).

\begin{figure}[t]
\begin{center}
\includegraphics[width=0.3\textwidth]{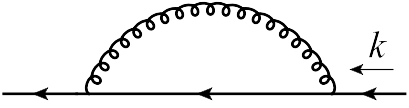}
\caption{The quark retarded self-energy $\varSigma^{R {\mathrm {(eq)}}}(k)$ whose momentum is hard in the leading order.
The solid line is the quark propagator and the cirly line is the gluon propagator.
We note that the gluon propagator in this figure is the HTL-resummed one~\cite{HTL-resum}, which results the anomalously large imaginary part ($\zetaf$) of $\varSigma^{R {\mathrm {(eq)}}}(k)$. 
}
\label{fig:quark-selfenergy}
\end{center}
\end{figure}

\begin{figure}[t]
\begin{center}
\includegraphics[width=0.4\textwidth]{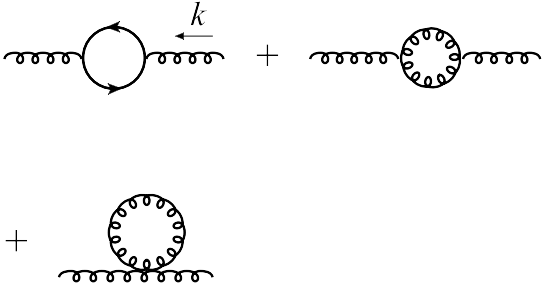} 
\caption{The gluon advanced self-energy $\varPi^{\A {\mathrm {(eq)}}ab\mu\nu}(k)$ whose momentum is hard in the leading order.
The notations are the same as Fig.~\ref{fig:quark-selfenergy}.
The gluon propagators in the second diagram are the HTL-resummed ones~\cite{HTL-resum}, which result the anomalously large imaginary part ($\zetab$) of $\varPi^{\A {\mathrm {(eq)}}ab\mu\nu}(k)$. 
The diagram which has ghost propagators is omitted since its contribution to the transverse sector of $\varPi^{\A {\mathrm {(eq)}}ab\mu\nu}(k)$ is zero.
}
\label{fig:gluon-selfenergy}
\end{center}
\end{figure}

Now let us see that the self-energy terms in Eq.~(\ref{eq:EOM-momentum}) is not negligible in comparison with the first term in the left-hand side, i. e., $2ik\cdot \partial_X K^a_i(k,X)$.
First, we evaluate the diagonal self-energy terms.
The dominant parts of the diagonal self-energies are those at the leading order and at equilibrium, whose diagrams are shown in Figs.~\ref{fig:quark-selfenergy} and \ref{fig:gluon-selfenergy} and whose expressions are given as follows: 
\begin{align}
\label{eq:selfenergy-fermion}
\{\Slash{k}, \varSigma^{\R {\text {(eq)}}}(k)\}&= \mf^2-2i\zetaf k^0,\\
\label{eq:selfenergy-boson}
P^T_{\alpha i}(k) \varPi^{\A {\text {(eq)}}ab\alpha\nu}(k)&= -\delta^{ab}P^{T\nu}_{i}(k)(\mb^2+2i\zetab k^0),
\end{align}
where $\mf\equiv \cp T\sqrt{\Cf}/2$ ($\mb\equiv \cp T\sqrt{(N+N_f/2)/6}$) is the quark (gluon) asymptotic thermal mass~\cite{thermal-mass}, with $\Cf\equiv (N^2-1)/(2N)$.
$\zeta_q$ ($\zeta_g$) is the quark (gluon) damping rate, which is of order $\cp^2T \ln (1/\cp)$~\cite{quark-gluon-damping}. 
We have used the fact that $k\sim T$ and $k^2=0$.
Since Eqs.~(\ref{eq:selfenergy-fermion}) and (\ref{eq:selfenergy-boson}) are of order $\cp^2T^2$, we confirm that the diagonal self-energy terms in the left-hand side of Eq.~(\ref{eq:EOM-momentum}) have the same order of magnitude as that of $2ik\cdot \partial_X K^a_i(k,X)$ in the left-hand side because $k\sim T$ and $\partial_X\sim\cp^2 T$.
For this reason, we can not neglect the diagonal self-energies in that equation, in contrast to the soft case ($\partial_X\sim \cp T$)~\cite{blaizot-HTL}.

\begin{figure}[t]
\begin{center}
\includegraphics[width=0.48\textwidth]{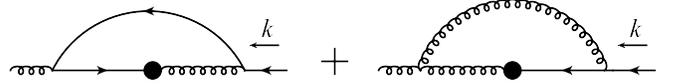} 
\caption{The off-diagonal self-energy $\VC^a_\mu(k, X)$ whose momentum is hard in the leading order.
The propagator that is composed of the solid line and the curly line with the black blob is the off-diagonal propagator. 
The other notations are the same as Fig.~\ref{fig:quark-selfenergy}.
}
\label{fig:vertex-correction}
\end{center}
\end{figure}

As in the diagonal self-energies case, the off-diagonal self-energy need to be taken into account.
The expression of the off-diagonal self-energy at the leading order, whose diagrams are shown in Fig.~\ref{fig:vertex-correction}, is given by the following equation:
\begin{align} 
\label{eq:selfenergy-offdiagonal} 
\begin{split}
&\VC^{\R a \mu}(k,X)=\\
&- \cp^2\int \frac{d^4 k'}{(2\pi)^4}\gamma^j t^b S^{0\R}(k+ k')\gamma^\mu t^a K^b_j(k',X) \\ 
 &~~~- \cp^2\int \frac{d^4 k'}{(2\pi)^4} \gamma^k t^d D^{0\R cd}_{lk}(k'-k)\\
 &~~~\times if^{abc}[g^{lj}(2k'-k)^\mu+g^{\mu j}(-k-k')^l\\
 &~~~+g^{\mu l}(2k-k')^j]
 K^{b}_j(k',X),
 \end{split}
\end{align}
where $S^{0\R}(k)\equiv-\Slash{k}/[(k_0+i\epsilon)^2-\vk^2]$ and $D^{0\R cd}_{lk}(k)\equiv -\delta^{cd}\{ P^T_{lk}(k)/[(k_0+i\epsilon)^2-\vk^2]+\hat{k}_l\hat{k}_k/(k_0+i\epsilon)^2\}$ are the free quark and gluon retarded propagator at equilibrium, respectively.
By making an order estimate of Eq.~(\ref{eq:selfenergy-offdiagonal}), we see that the off-diagonal self-energy term in Eq.~(\ref{eq:EOM-momentum}) has the same order of magnitude as that of $2ik\cdot \partial_X K^a_i(k,X)$, and thus that term can not be neglected.

\begin{figure*}[t]
\begin{center} 
\includegraphics[width=0.8\textwidth]{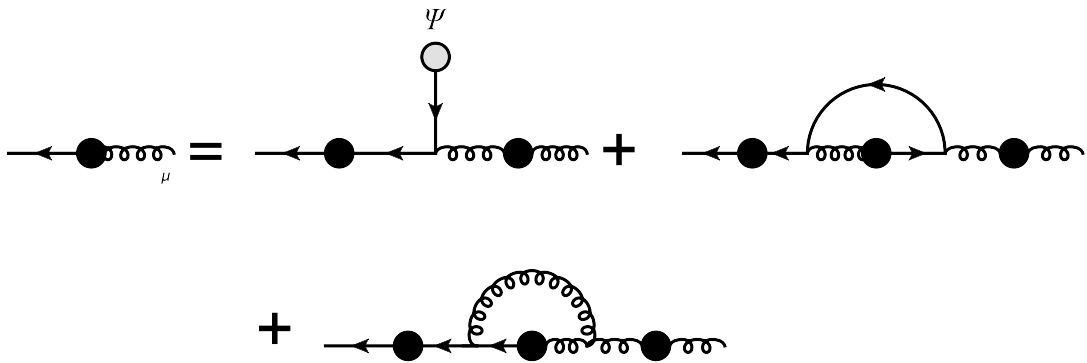}
\caption{The diagrammatic representation of self-consistent equation for $K^a_i$ at the leading order.
For simplicity, $A^a_\mu$ is set to be zero.
The solid (curly) line with a black blob is the resummed quark (gluon) propagator, whose information is reflected in Eqs.~(\ref{eq:selfenergy-fermion}) and (\ref{eq:selfenergy-boson}). 
}
\label{fig:K}
\end{center}
\end{figure*}

By inserting Eqs.~(\ref{eq:freepropagator-fermion})$\sim$(\ref{eq:selfenergy-offdiagonal}) into Eq.~(\ref{eq:EOM-momentum}), we obtain the following equation (see Appendix \ref{app:off-diagonal} for a detailed derivation): 
\begin{align} 
\label{eq:EOM-result}
\begin{split}
&\left(2ik\cdot \partial_X-2\cp k\cdot A^b(X)t^b+2i\zeta k^0+\delta m^2 \right)K^a_i(k,X)\\
&-2\cp k\cdot A^c(T^c)_{ab}K^b_i(k,X)\\
&= \Slash{k}\rho^0(k)P^T_{ij}(k)(\nb(k^0) + \nf(k^0))\\
&\times\cp t^{a}\left(\gamma^{j}\varPsi(X)+\cp t^b\int \frac{d^4 k'}{(2\pi)^4} \frac{k^k\gamma^j+\gamma^kk'{}^{j}}{k\cdot k'}  K^b_k(k',X)\right),
\end{split}
\end{align}
where we have introduced 
\begin{align}
\delta m^2\equiv \mb^2-\mf^2=\cp^2T^2\left(\frac{N}{24}+\frac{1}{8N}+\frac{N_f}{12}\right)
\end{align}
 and $\zeta\equiv\zetaf+\zetab$, and used the relation $P^{T}_{ik}(k) K^{a k}(k,X)\simeq -K^{a}_{i}(k,X)$ again.
We note that this equation transforms in a covariant way under the gauge transformation Eq.~(\ref{eq:gauge-transform}), which can be confirmed by using the transformation property of $K^a_i(k,X)$~\cite{blaizot-review}, 
\begin{align}
K^a_i(k,X)\rightarrow h(X)K^b_i(k,X) \tilde{h}^{ab}(X) ,
 \end{align}
 where $\tilde{h}(x)\equiv \exp[i\theta^a(x)T^a]$.
The diagram that corresponds to Eq.~(\ref{eq:EOM-result}) is shown in Fig.~\ref{fig:K}.
From Eq.~(\ref{eq:EOM-result}), we get the following equation for $K_i(k,X)\equiv t^aK^a_i(k,X)$:
\begin{align} 
\label{eq:EOM-result2}
\begin{split}
&\left(2ik\cdot \partial_X-2\cp k\cdot A^a(X)t^a+2i\zeta k^0+\delta m^2 \right)K_i(k,X)\\
&= \cp \Cf\Slash{k}\rho^0(k)P^T_{ij}(k)(\nb(k^0) + \nf(k^0))\\
&~~~\times\left(\gamma^{j}\varPsi(X)+\cp \int \frac{d^4 k'}{(2\pi)^4} \frac{k^k\gamma^j+\gamma^kk'{}^{j}}{k\cdot k'}  K_k(k',X)\right).
\end{split}
\end{align}
From this equation, we confirm that $K_i(k,X)$ transforms under the gauge transformation Eq.~(\ref{eq:gauge-transform}) in the same way as $\varPsi(X)$ in contrast to $K^a_i(k,X)$. 

By introducing $\varLambda_{\mu\pm}(\vk,X)$, which is defined by $K_\mu(k,X)\equiv2\pi\delta(k^2)[\theta(k^0)\varLambda_{\mu +}(\vk,X)+\theta(-k^0)\varLambda_{\mu -}(-\vk,X)]$, we arrive at the following GBE\footnote{In the case of the analysis of the lightlike momentum instead of the ultrasoft one, a equation which is similar to this equation was obtained before~\cite{markov}. 
However, the equation in Ref.~\cite{markov} does not contain all the leading contributions:
The equation does not contain the third term in the left-hand side and the second term in the right-hand side in Eq.~(\ref{eq:kinetic-eq}).}: 
\begin{align} 
\label{eq:kinetic-eq}
\begin{split} 
&\left(2iv\cdot D_X[A]\pm\frac{\delta m^2}{|\vk|}+2i\zeta\right)\Slash{\varLambda}_{\pm}(\vk, X)\\
&\quad=2\cp\Cf \Slash{v}[\nb(|\vk|)+\nf(|\vk|)]\varPsi(X)\\
&\qquad-\cp^2\Cf\gamma_{i}\Slash{v}[\nf(|\vk|)+\nb(|\vk|)]P^{\nu i }_{\mathrm T}(v)\\
&\qquad\times\sum_{s=\pm}\int\frac{d^3\vk '}{(2\pi)^3}\frac{1}{2|\vk '|}\frac{s|\vk|v^\alpha\gamma_\nu\pm |\vk'|v_{ \nu}'\gamma^\alpha}{|\vk| |\vk '|v\cdot v'}\varLambda_{s \alpha}(\vk',X),
\end{split} 
\end{align}
where $v^\mu\equiv (1, \hat{\vk})$ and $v'{}^\mu\equiv (1, \hat{\vk}')$, with $\hat{\vk}\equiv \vk/|\vk|$.
Since the structure of this equation is the same as that in QED~\cite{kinetic-offdiagonal}, the interpretation of each term in that equation is the same as in the QED case~\cite{kinetic-offdiagonal}.

%%%%%%%%%%%%%%%%%%%%%%%%%%%%%%%%
\subsection{Linear response regime}
\label{ssc:linear-response}

When $A^\mu_a=0$, the color structure of the off-diagonal propagator becomes simple.
In this case, Eq.~(\ref{eq:EOM-result}) becomes
\begin{align}
\label{eq:EOM-result-linear-response}
\begin{split}
&\left(2ik\cdot \partial_X+2i\zeta k^0+\delta m^2 \right)K^a_i(k,X)\\
&= \Slash{k}\rho^0(k)P^T_{ij}(k)(\nb(k^0) + \nf(k^0))\\
&\times\cp t^{a}\left(\gamma^{j}\varPsi(X)+\cp t^b\int \frac{d^4 k'}{(2\pi)^4} \frac{k^k\gamma^j+\gamma^kk'{}^{j}}{k\cdot k'}  K^b_k(k',X)\right).
\end{split}
\end{align}
This equation tells us that $K^a_i$ has the following color structure:
\begin{align}
\label{eq:color-structure-linear}
K^a_i(k,X)=t^a \frac{K_i(k,X)}{\Cf}.
\end{align}
In the present case, $K_i(k,X)$ has the same color structure as that of $\varPsi(X)$.
For later use, we point out that $\VC^{\R a}_i$ has also the same color structure as $K^a_i$.
By using Eq.~(\ref{eq:color-structure-linear}), Eq.~(\ref{eq:EOM-result-linear-response}) becomes
\begin{align}
\label{eq:EOM-result-linear-response-color}
\begin{split}
&\left(2ik\cdot \partial_X+2i\zeta k^0+\delta m^2 \right)K_i(k,X)\\
&= \cp\Cf\Slash{k}\rho^0(k)P^T_{ij}(k)(\nb(k^0) + \nf(k^0))\\
&\times \left(\gamma^{j}\varPsi(X)
+\cp \int \frac{d^4 k'}{(2\pi)^4} \frac{k^k\gamma^j+\gamma^kk'{}^{j}}{k\cdot k'}  K_k(k',X)\right).
\end{split}
\end{align} 

%%%%%%%%%%%%%%%%%%%%%%%%%%%%%%%%
\subsection{Equivalence between generalized Boltzmann equation and resummed perturbation}
\label{ssc:equivalence}

Let us show that the GBE in $A^a_\mu(X)=0$ case is equivalent to the self-consistent equation derived in the resummed perturbation~\cite{lebedev}.
For this purpose, we write the off-diagonal propagator in terms of the off-diagonal self-energy using Eq.~(\ref{eq:EOM-result-linear-response-color}):
\begin{align}
\label{eq:K-vertex}
\begin{split} 
&\left(2k\cdot p+2i\zeta k^0+\delta m^2 \right)K_i(k,p)\\
&= \cp\Slash{k}\rho^0(k)P^T_{ij}(k)(\nb(k^0) + \nf(k^0))
\tilde{\varGamma}^{j}(k,p).
\end{split}
\end{align}
Here we have performed the Fourier transformation as $f(p)\equiv \int d^4 X/(2\pi)^4 e^{ip\cdot X} f(X)$, with $f(X)$ is an arbitrary function, and introduced $\tilde{\varGamma}^{j}(k,p)$, which is defined as $\tilde{\varGamma}^{aj}(k,p)=t^a\tilde{\varGamma}^{j}(k,p)/\Cf$.
Since $\VC^{\R}_i$ has the same color structure as that of $\varPsi$, so does $\tilde{\varGamma}^{j}$.
Now we can obtain the self-consistent equation in terms of the off-diagonal self-energy, by inserting Eq.~(\ref{eq:K-vertex}) into Eq.~(\ref{eq:app-offdiagonal-result}):
\begin{align}
\label{eq:BS}
\begin{split}
& \frac{{\varGamma}^{i}(k,p)}{\Cf}
\simeq \gamma^i +  \cp^2 \int \frac{d^4 k'}{(2\pi)^4} \frac{k^j\gamma^i+\gamma^jk'{}^{i}}{k\cdot k'}  \\
&~~~\times\frac{\Slash{k'}\rho^0(k')P^T_{jk}(k')(\nb(k'{}^0) + \nf(k'{}^0))}
{2k'\cdot p+2i\zeta k'{}^0+\delta m^2} {\varGamma}^{k}(k',p),
\end{split}
\end{align}
where we have introduced $\varGamma_i(k,p)$ by $\V_i(k,p) \equiv \varGamma_i(k,p)\varPsi(p)$; note that $\varGamma_i(k,p)$ does not have a color structure.
This equation is none other than the self-consistent equation in Ref.~\cite{lebedev}.
By comparing Eq.~(\ref{eq:BS}) with the self-consistent equation in Ref.~\cite{lebedev}, we see that $\varGamma^a_i(k,p)\equiv t^a\varGamma_i(k,p)/\Cf$ is the vertex function introduced in Ref.~\cite{ultrasoft-QED, lebedev} whose momenta are hard and ultrasoft.
We see that the color structure of $\varGamma^a_i$ is the same as that of the bare vertex, $t^a\gamma_i$, as was shown in Ref.~\cite{lebedev} in a diagrammatic way, by using the fact that $\varGamma_i$ does not have a color structure.

Also the induced fermion source can be expressed in terms of the vertex correction:
From the definition of $\etaind(X)$ in Eq.~(\ref{eq:induced-current-fermion}) and its Fourier transformation, we have
\begin{align} 
\label{eq:etaind-K}
\begin{split}
\etaind(p)&= \cp t^a \gamma^\mu \int\frac{d^4k}{(2\pi)^4}K^a_\mu(k,p) \\
&\simeq \cp \gamma^i \int\frac{d^4k}{(2\pi)^4}K_i(k,p).
\end{split}
\end{align}
By using the relation~\cite{kinetic-offdiagonal, blaizot-HTL,blaizot-review} 
\begin{align}
\label{eq:selfenergy-current}
\etaind(p)=\varSigma^\R(p)\varPsi(p), 
\end{align}
which is valid in the linear response regime, and Eq.~(\ref{eq:K-vertex}), we obtain
\begin{align} 
\begin{split}
\varSigma^\R(p) &= \cp^2 \int \frac{d^4k}{(2\pi)^4}\gamma^i  \frac{\Slash{k}\rho^0(k)P^T_{ij}(k)(\nb(k^0) + \nf(k^0))}{2k\cdot p+2i\zeta k^0+\delta m^2 }\\
&~~~\times{\varGamma}^{j}(k,p).
\end{split}
\end{align}
This expression coincides with that in Ref.~\cite{lebedev}.
Thus we see that our kinetic equation is equivalent to the self-consistent equation in the resummed perturbation~\cite{lebedev}.
This equivalence establishes the foundation of the resummed perturbation scheme.

By using the correspondence between the self-consistent equation in the resummed perturbation theory~\cite{lebedev, ultrasoft-QED} and the GBE (Eq.~(\ref{eq:kinetic-eq})), we can obtain kinetic interpretations of the procedures of the resummed perturbation theory.
Since the structure of the GBE is almost the same as that in the QED case~\cite{kinetic-offdiagonal}, we refrain from giving the interpretation of the resummation scheme in this paper.

%%%%%%%%%%%%%%%%%%%%%%%%%%%%%%%%
\subsection{Ward-Takahashi identity}

We show explicitly that the Ward-Takahashi (WT) identity derived from $SU(N)$ gauge symmetry is satisfied in the present analysis in this subsection.
We also show that the WT identity can be derived from the conservation law of the current.

By multiplying Eq.~(\ref{eq:selfenergy-offdiagonal}) by $k^\mu$, we obtain
\begin{align}
\label{eq:WT-calculation}
 k_\mu\VC^{\R a \mu}(k,X)&= \cp t^a\etaind(X).
\end{align}
See Appendix~\ref{app:off-diagonal} for the detailed calculation.
This equation generates the WT identity at the leading order by setting $A^a_\mu=0$.

Now we show that the WT identity can be derived from the conservation law of the color current, which reads~\cite{blaizot-HTL, blaizot-review}
\begin{align}  
\label{eq:current-conservation}
\partial_\mu \jind^\mu(x)&=i\cp t^a(\overline{\varPsi}(x)t^a\etaind(x)-\overline{\eta}_{\text{ind}}(x)t^a\varPsi(x)),
\end{align}
where we have set $A^a_\mu=0$ and introduced $\jind^\mu\equiv t^a \jind^{a\mu}$.
By differentiating Eq.~(\ref{eq:current-conservation}) with respect to $\overline{\varPsi}(y)$, we get
\begin{align}
\partial_{\mu x} \VC^{ \mu}_{}(y,x)
&= \cp t^a\delta(x^0-y^0)\delta^{(3)}(\vx-\vy)t^a\etaind(x).
\end{align} 
Here we have set $x^0$, $y^0\in C^+$, and neglected the term which is of order $\cp^3 T\varPsi$.
By multiplying this equation by $\int d^4s \exp({ik\cdot s})$ and taking only the leading-order terms~\cite{kinetic-offdiagonal}, we find 
\begin{align}
\label{eq:WT2}
-k_\mu \VC^{\R\mu}(-k,X)&= \cp t^a t^a \etaind(X).
\end{align} 
We see that this equation is nothing but Eq.~(\ref{eq:WT-calculation}) by using the color structure of $\VC^{\R\mu}$ in the linear response regime.

%%%%%%%%%%%%%%%%%%%%%%%%%%%%%%%%%%%%%%%%%%%%%%%%%%%%%%%%%%%%%%%%%%
\section{Ultrasoft fermion mode}
\label{sec:fermion-mode}
In this section, we show the existence of the ultrasoft mode in QCD, and obtain the expressions for the pole position and the residue of that mode by solving the self-consistent equation that determines the off-diagonal propagator in the linear response regime, Eq.~(\ref{eq:EOM-result-linear-response-color}).
We focus on the momentum region $\tilde{p}\ll \cp^2T$ with $\tilde{p}\equiv (p^0+i\zeta, \vp)$, in which we can solve the self-consistent equation analytically.

\begin{figure}[t]
\begin{center} 
\includegraphics[width=0.45\textwidth]{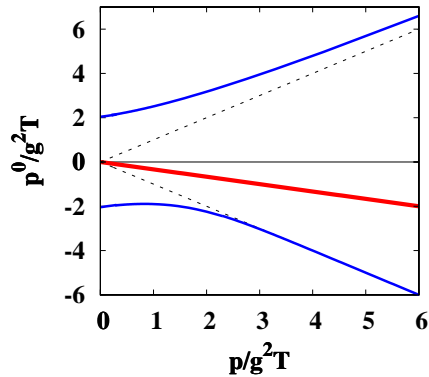}
\caption{The dispersion relation in the fermionic sector.
The vertical axis is the energy $p^0$, while the horizontal axis is the momentum $|\vp|$. 
The solid (blue) lines corresponds to the normal fermion and the antiplasmino while the bold solid (red) one to the ultrasoft mode.
The dotted lines denote the light cone.
Since our analysis on the ultrasoft mode is valid only for $|\vp|\ll \cp^2T$, the plot for $|\vp|\gtrsim \cp^2T$ may not have a physical meaning.
The residue of the antiplasmino becomes exponentially small for $|\vp|\gg \cp T$, 
so the plot of the antiplasmino does not represent physical excitation for $|\vp|\gg \cp T$, either.
The parameters are set as follows:
$\cp=0.2$ and $N=3$.}
\label{fig:dispersion}
\end{center}
\end{figure}

As a result of the analysis using Eqs.~(\ref{eq:EOM-result-linear-response-color}) and (\ref{eq:etaind-K}), we find that the quark retarded self-energy with the momentum that satisfies $\tilde{p}\ll \cp^2 T$ has the following expression:
\begin{align}
\label{eq:fermion-mode-result}
\varSigma^\R(p) &= -\frac{Y(\tilde{p})}{Z},
\end{align} 
where $Y(\tilde{p})\equiv \tilde{p}{}^0\gamma^0+\hat{\vp}\cdot\vgamma/3$ and $Z\equiv \cp^2\Cf/(16\pi^2\lambda^2A^2)$, with $\lambda\equiv \cp^2T^2\Cf/(8\delta m^2)$ and the expression of $A$ is obtained in Eq.~(\ref{eq:A-result}):
See Appendix~\ref{app:ultrasoft-mode} for a detailed derivation.
Since $Z\sim \cp^2$, $\varSigma^\R(p)$ is much larger than the inverse of the free-part of the quark propagator.
Thus, the ultrasoft quark retarded propagator at the leading order is
\begin{align} 
\begin{split}
S^\R(p)&\simeq \frac{1}{\varSigma^\R(p)} \\
&= -\frac{Z}{2}\left(\frac{\gamma^0-\hat{\vp}\cdot\vgamma}{p^0+|\vp|/3+i\zeta}
+\frac{\gamma^0+\hat{\vp}\cdot\vgamma}{p^0-|\vp|/3+i\zeta} \right).
\end{split}
\end{align}
This expression implies that there is a fermionic excitation in the ultrasoft region, which was suggested originally in Ref.~\cite{lebedev}, whose pole position is 
\begin{align}
p^0= \mp\frac{|\vp|}{3}-i\zeta.
\end{align}
This expression means that that excitation's velocity is $1/3$ and damping rate is $\zeta\sim \cp^2T\ln(1/\cp)$.
The dispersion relation of the mode is plotted in Fig.~\ref{fig:dispersion} with those of the two soft fermionic excitations~\cite{plasmino}, for comparison.
The residue of that pole is 
\begin{align} 
\begin{split}
Z&= \frac{\cp^2 }{16\pi^2\Cf}\left(\Cf+\frac{8\delta m^2}{\cp^2T^2}\right)^2 \\
 &=\cp^2 \frac{N}{8\pi^2(N^2-1)}\left(\frac{5}{6}N+\frac{1}{2N}+\frac{2}{3}N_f \right)^2.
 \end{split}
\end{align} 
If we set $N=3$ and $N_f=3$, the expression becomes
\begin{align}
Z&=  \cp^2\frac{49}{48\pi^2}.
\end{align}
We emphasize that these expressions are obtained in this paper for the first time in QCD. 
These results are summarized in Table~\ref{tab:properties}.
We note that the expressions are qualitatively the same as those in the Yukawa model and QED~\cite{ultrasoft-QED}.

\begin{table}[t]
\begin{center}
\caption{The dispersion relation, the damping rate, and the residue of the ultrasoft fermion mode in QCD.}
\begin{tabular}{c|c}
\hline
dispersion relation &  $\mp |\vp|/3$ \\ 
damping rate &  $\zeta_f+\zeta_b \sim \cp^2 T\ln \cp^{-1}$\\
residue &  $\cp^2 \frac{N}{8\pi^2(N^2-1)}\left(\frac{5}{6}N+\frac{1}{2N}+\frac{2}{3}N_f \right)^2 \sim \cp^2$ \\ 
\hline
\end{tabular}
\label{tab:properties}
\end{center}
\end{table}

%%%%%%%%%%%%%%%%%%%%%%%%%%%%%%%%
\section{$n$-point function ($n\geq 3$)}
\label{sec:higher}
In this section, we obtain the equation determining the $n$-point function whose external lines consist of a pair of quarks and $(n-2)$ gluons with ultrasoft momenta, where $n\geq 3$. 
We also make an order estimate of the $n$-point function by using that equation.
We note that the equation was obtained by going beyond the linear response theory, and it is a merit of the Kadanoff-Baym formalism.

The fermion induced source also generates $n$-point function not only two point function of the quark, which was analyzed in Sec.~\ref{ssc:linear-response}.
Here we work in the $n=3$ case.
In this case, we use the following relation~\cite{blaizot-HTL,blaizot-ultrasoft,kinetic-offdiagonal}:
\begin{align}
\label{eq:higher-vertex}
\begin{split}
&\delta^{(4)}(p-q-r)\cp\delta\varGamma^a_{\mu}(p,-q,-r)\\
&= \left. \frac{\delta^2 \etaind (p)}{\delta\varPsi(q)\delta A^{a\mu}(r)}\right|_{A, \varPsi=0} \\
&\simeq\cp \gamma^i \int \frac{d^4k}{(2\pi)^4}  \left. \frac{\delta^2 K_i (k,p)}{\delta\varPsi(q)\delta A^{a\mu}(r)}\right|_{A, \varPsi=0}
\end{split}
\end{align}
Here $\cp\delta\varGamma^a_{\mu}(p,-q,-r)$ is the quark-gluon vertex correction function whose momenta ($p$, $-q$, and $-r$) are ultrasoft.
To determine $\delta\varGamma^a_{\mu}(p,-q,-r)$, we expand the off-diagonal propagator in the following way:
\begin{align}
K_i(k,X)&= K_i(k,X)_{A=0}+\delta K_i(k,X)+O(A^2\varPsi),
\end{align}
where $K_i(k,X)_{A=0}$ is in the linear order in $\varPsi$ while $\delta K_i(k,X)$ is so in $\varPsi$ and $A^a_\mu$.
By collecting the terms that are in the linear order in $\varPsi$ and $A^a_\mu$, Eq.~(\ref{eq:EOM-result2}) becomes
\begin{align}
\label{eq:final-3point}
\begin{split}
&\left(2ik\cdot \partial_X+2i\zeta k^0+\delta m^2 \right)\delta K_i(k,X)\\
&~~~-2\cp k\cdot A^a(X)t^aK_i(k,X)_{A=0}\\
&= \cp^2 \Cf \Slash{k}\rho^0(k)P^T_{ij}(k)(\nb(k^0) + \nf(k^0))\\
&~~~\times\int \frac{d^4 k'}{(2\pi)^4} \frac{k^k\gamma^j+\gamma^kk'{}^{j}}{k\cdot k'}  \delta K_k(k',X).
\end{split}
\end{align}
Since $K_i(k,X)_{A=0}$ is determined from Eq.~(\ref{eq:EOM-result-linear-response-color}), we can determine $\delta K_i(k,X)$ from this equation.
By using $\delta K_i(k,X)$, $\delta\varGamma^a_{\mu}(p,-q,-r)$ can be obtained from Eq.~(\ref{eq:higher-vertex}).
The similar analysis can be done also in the case of $n\geq 4$.

Here we make an order estimate of the three point function.
Since $k\sim T$ and $\partial_X\sim \cp^2T$, we see that $K^i_{A=0}\sim \cp^{-1}T^{-3}\varPsi$ by using Eq.~(\ref{eq:EOM-result-linear-response-color}).
From this estimate and Eq.~(\ref{eq:final-3point}), we find that $\delta K_i\sim \cp^{-2} T^{-4}\varPsi A^\mu$.
Therefore, the vertex correction whose momenta are ultrasoft is estimated as $\cp\delta\varGamma^\mu\sim \cp^{-1}$.
We note that this quantity is much larger than the bare vertex, which is of order $\cp$.
Similarly, in the case that $n\geq 4$, the $n$-point function is of order $\cp^{2-n}$.

%%%%%%%%%%%%%%%%%%%%%%%%%%%%%%%%%%%%%%%%%%%%%%%%%%%%%%%%%%%%%%%%%%
\section{Summary and Concluding Remarks} 
\label{sec:summary}

We have derived the generalized Boltzmann equation from the Kadanoff-Baym equation for quark excitation with an ultrasoft momentum ($\sim\cp^2T$) near equilibrium, and showed that the equation is equivalent to the self-consistent equation obtained in the resummation scheme~\cite{lebedev} that is used in the analysis of the quark propagator with an ultrasoft momentum, in the linear response regime in QCD at extremely high $T$.
We have obtained the expression of the dispersion relation, the damping rate, and the strength of the ultrasoft fermion mode for the first time in QCD, by solving the generalized and linearized Boltzmann equation.
We have showed that the Ward-Takahashi identity is satisfied in our approach, and that identity can be derived from the conservation law of the current.
We also have obtained the equation that determines the $n$-point function containing a pair of quarks and $(n-2)$ gluon external lines whose momenta are ultrasoft.
We note that this equation was obtained by retaining the nonlinear response caused by the gluon average field in the Kadanoff-Baym equation, and the resummed perturbation has not been able to produce the equation.

As we mentioned in Sec.~\ref{sec:background-gauge}, the possible gauge-fixing dependence\footnote{We note that the analysis in the Coulomb gauge and the temporal gauge was performed in QED, and we found no differences  in the ultrasoft fermion propagator in the two gauge-fixing~\cite{kinetic-offdiagonal}.} should be checked by adopting gauge-fixing which is other than the temporal gauge. 
Also, though the analyses in this paper and the previous one~\cite{ultrasoft-QED} are restricted to the Yukawa model, QED, and QCD, it is possible to extend the analysis to other fermion-boson system.
Especially, investigating the fermionic spectrum with ultrasoft momentum at high temperature in the Weinberg-Salam theory is an interesting task since the investigation can be relevant to the analysis of the properties of the early universe~\cite{neutrino}.

%%%%%%%%%%%%%%%%%%%%%%%%%%%%%%%%%%%%%%%%%%%%%%%%%%%%%%%%%%%%%%%%%%
\section*{Acknowledgement}

This work was supported by the Grant-in-Aid for the Global COE Program ``The Next Generation of Physics, Spun from Universality and Emergence'' from the Ministry of Education, Culture, Sports, Science and Technology (MEXT) of Japan.
The author thanks T.~Kunihiro for his careful reading of the manuscript.
The author also thanks J.~P.~Blaizot, Y.~Hidaka, and E.~Iancu for fruitful discussions and comments.
This work was supported by JSPS KAKENHI Grant Numbers 24$\cdot$56384.

%%%%%%%%%%%%%%%%%%%%%%%%%%%%%%%%%%%%%%%%%%%%%%%%%%%%%%%%%%%%%%%%%%%
\appendix
\section{SMALLNESS OF LONGITUDINAL COMPONENT OF OFF-DIAGONAL PROPAGATOR}
\label{app:longitudinal}

In this Appendix, we show that the longitudinal component of the off-diagonal propagator, $k^i K^a_i(k,X)$, is negligible compared with the transverse component.

By multiplying Eq.~(\ref{eq:EOM-wigner-fermion}) by $(\Slash{k}+i\Slash{\partial}_X/2-\varSigma^\R(k,X))$ and subtracting Eq.~(\ref{eq:EOM-wigner-boson}) from that quantity, we get 
\begin{align}
\begin{split}
&  -k^i k^j  K^{a}_{j}(k,X) 
=  -\varPi^{\A abij}_{}(k,X)K^{b}_{j}(k,X)\\
&~~~+\cp( S^<(k,X)\V^{ai}(k,X)+\Slash{k}\V^{ bj}(k,X)D^{< ba}_{ji}(k,X))  .
\end{split}
\end{align} 
Here we have set $A^a_\mu(X)=0$ for simplicity, neglected the terms that are of order $\cp^2T^2 K^a_i(k,X)$, and set $k\rightarrow i$ in Eq.~(\ref{eq:EOM-wigner-boson}).
From this equation and Eq.~(\ref{eq:EOM-momentum}), we see that the longitudinal component of the off-diagonal propagator, $\hat{k}^iK^a_i(k,X)$, has the same order of magnitude as $\cp^2 K^a_i(k,X)$.

%%%%%%%%%%%%%%%%%%%%%%%%%%%%%%%%%%%%%%%%%%%%%%%%%%%%%%%%%%%%%%%%%%%
\section{CALCULATION OF OFF-DIAGONAL SELF-ENERGY} 
\label{app:off-diagonal}

We derive Eqs.~(\ref{eq:EOM-result}) and (\ref{eq:WT-calculation}) in this Appendix.
First we derive Eq.~(\ref{eq:EOM-result}).
To this end, we transform Eq.~(\ref{eq:selfenergy-offdiagonal}) in the following way:
\begin{align}
\begin{split} 
&\VC^{\R a i}(k,X)\simeq \\
&~~~ \cp^2\int \frac{d^4 k'}{(2\pi)^4} \frac{k^j\gamma^i+\gamma^jk'{}^{i}}{k\cdot k'} t^b t^a K^b_j(k',X) \\ 
 &~~~+  if^{abc}t^c\cp^2\int \frac{d^4 k'}{(2\pi)^4} \gamma^k  \\
 &~~~\times\left(\frac{P^T_{lk}(k'-k)}{-2k\cdot k'}+\frac{(k'-k)_l(k'-k)_k}{(k'{}^0-k^0)^2|\vk'-\vk|^2}\right)\\
 &~~~\times[2k'{}^ig^{lj}-g^{ij}(k+k')^l+2k^jg^{il} ]
 K^{b}_j(k',X) .
 \end{split}
\end{align}
Here we have used $k^iK^a_i(k,X)\simeq 0$ and $k^2\simeq k'{}^2\simeq 0$, and neglected the term that is proportional to $k^i$ since this term does not contribute to Eq.~(\ref{eq:EOM-result}) due to the presence of $P^T_{ij}(k)$. 
We also have used $\Slash{k}K^a_i(k,X)\simeq 0$, which can be checked by multiplying Eq.~(\ref{eq:EOM-momentum}) by $\Slash{k}$ from the left.
Using $k^2\simeq k'{}^2\simeq 0$ again, we get
\begin{align} 
\label{eq:app-offdiagonal-B}
\begin{split} 
&\VC^{\R a i}(k,X)\simeq \\
&~~~\cp^2(t^at^b -if^{abc}t^c)\int \frac{d^4 k'}{(2\pi)^4} \frac{k^j\gamma^i+\gamma^jk'{}^{i}}{k\cdot k'}  K^b_j(k',X) \\ 
 &~~~ -if^{abc}t^c\cp^2\int \frac{d^4 k'}{(2\pi)^4} \frac{\gamma^k}{2k\cdot k'}  
 \left(\delta_{lk}-\frac{(k'-k)_l(k'-k)_k}{(k'{}^0-k^0)^2}\right)\\
 &~~~\times[2k'{}^ig^{lj}-g^{ij}(k+k')^l+2k^jg^{il} ] K^{b}_j(k',X) \\
&\simeq \cp^2(t^at^b -if^{abc}t^c)\int \frac{d^4 k'}{(2\pi)^4} \frac{k^j\gamma^i+\gamma^jk'{}^{i}}{k\cdot k'}  K^b_j(k',X) \\ 
 &~~~ -if^{abc}t^c\cp^2\int \frac{d^4 k'}{(2\pi)^4} \frac{\gamma^k}{k\cdot k'}  
 \Bigl[k'{}^ig^{kj}+k^jg^{ik}\\
 &~~~ -g^{ij}\frac{k^k(|\vk'|^2-k^0k'{}^0)+k'{}^k(|\vk|^2-k^0k'{}^0)}{(k'{}^0-k^0)^2}\Bigr]
 K^{b}_j(k',X) . 
\end{split}
\end{align}
In the last line we have used $k^2\simeq k'{}^2\simeq 0$ and $k^iK^a_i(k,X)\simeq 0$, and neglected the term that is proportional to $k^i$ as before.
We note that in the derivation of Eq.~(\ref{eq:EOM-result}), the term that is proportional to $\Slash{k}$ in Eq.~(\ref{eq:app-offdiagonal-B}) can be neglected since it yields higher order contribution after being multiplied by $\Slash{k}$.
By neglecting the term that is proportional to $\Slash{k}$ and using $\Slash{k}K^a_i(k,X)\simeq 0$, we see that $\gamma^k k^k\simeq \gamma^0 k^0$ and $\gamma^k k'{}^k\simeq \gamma^0 k'{}^0$ in Eq.~(\ref{eq:app-offdiagonal-B}).
Using these properties, it is easily shown that the $g^{ij}$ term in Eq.~(\ref{eq:app-offdiagonal-B}) vanishes.
Thus that equation becomes
\begin{align}
\label{eq:app-offdiagonal-result}
\begin{split}
&\VC^{\R a i}(k,X)\simeq  \cp^2 t^at^b\int \frac{d^4 k'}{(2\pi)^4} \frac{k^j\gamma^i+\gamma^jk'{}^{i}}{k\cdot k'}  K^b_j(k',X) . 
\end{split}
\end{align}
By substituting this equation into Eq.~(\ref{eq:EOM-momentum}), we will get Eq.~(\ref{eq:EOM-result}).

Next, we derive Eq.~(\ref{eq:WT-calculation}).
By multiplying Eq.~(\ref{eq:selfenergy-offdiagonal}) by $k_\mu$, we get
\begin{align}
\begin{split} 
& k_\mu\VC^{\R a \mu}(k,X) \\
 &= \cp^2\int \frac{d^4 k'}{(2\pi)^4}\gamma^j t^b t^a \frac{\Slash{k}+\Slash{k}'}{(k+k')^2}(\Slash{k}+\Slash{k}'-\Slash{k}')K^b_j(k',X) \\ 
 &~~~ + if^{abc}\cp^2\int \frac{d^4 k'}{(2\pi)^4} \gamma^k t^c \\
 &~~~\times\left(\frac{P^T_{lk}(k'-k)}{(k'-k)^2}+\frac{(k'-k)_l(k'-k)_k}{(k'{}^0-k^0)^2|\vk '-\vk|^2}\right)\\ 
 &~~~\times [g^{lj}(2k'\cdot k-k^2)-k^j(k+k')^l+k^l(2k-k')^j] \\
 &~~~\times K^{b}_j(k',X) \\
 &\simeq \cp^2\int \frac{d^4 k'}{(2\pi)^4} (t^a t^b-if^{abc}t^c )\gamma^j K^b_j(k',X) \\  
 &~~~ +\cp^2\int \frac{d^4 k'}{(2\pi)^4} \frac{ if^{abc}t^c\gamma^k}{(k'-k)^2}  
 \Bigl( \delta_{lk}
- \frac{(k'-k)_l(k'-k)_k }{(k'{}^0-k^0)^2}\Bigr) \\
 &~~~\times \left[g^{lj}(2k'\cdot k-k^2)+k^j(k-k')^l \right]K^{b}_j(k',X) . 
 \end{split}
\end{align}
Here we have used $\Slash{k}'K^i(k',X)\simeq 0$ and $k'{}^iK^a_i(k',X)\simeq 0$ in the last line.  
Using $k'{}^2\simeq 0$ and $k'{}^iK^a_i(k',X)\simeq 0$ again, we get
\begin{align}
\begin{split}
& k_\mu\VC^{\R a \mu}(k,X) \\
&\simeq \cp t^a\etaind(X) -\cp^2 if^{abc}t^c\int \frac{d^4 k'}{(2\pi)^4} \gamma^j K^b_j(k',X) \\  
 &~~~ + if^{abc}t^c\cp^2\int \frac{d^4 k'}{(2\pi)^4} \gamma^k \Bigl[ \delta_{jk}
 +\frac{(k'-k)_k k^j}{(k'-k)^2(k'{}^0-k^0)^2}\\ 
 &~~~\times\left((k'{}^0-k^0)^2-(k'-k)^2-|\vk'-\vk|^2 \right)\Bigr] K^{b}_j(k',X)\\
&= \cp t^a\etaind(X) . 
 \end{split}
\end{align}
We note that the contribution from the gluon self-interaction partly cancels the contribution from the quark-gluon interaction.

%%%%%%%%%%%%%%%%%%%%%%%%%%%%%%%%%%%%%%%%%%%%%%%%%%%%%%%%%%%%%%%%%%%
\section{CALCULATION OF RETARDED QUARK SELF-ENERGY WITH ULTRASOFT MOMENTUM} 
\label{app:ultrasoft-mode}

We derive Eq.~(\ref{eq:fermion-mode-result}) in this Appendix.
First, we expand the off-diagonal propagator as 
\begin{align} 
K_i(k,p)&=\Kzero_i(k,p)+\Kone_i(k,p). 
\end{align}
Here $\Kzero_i(k,p)$ ($\Kone_i(k,p)$) is the zeroth (first) order term of the expansion in terms of $\tilde{p}/(\cp^2 T)$.
By collecting the zeroth order terms in Eq.~(\ref{eq:EOM-result-linear-response-color}) after the Fourier transformation, we get
\begin{align}
\label{eq:app-ultrasoft-1}
\begin{split} 
&\delta m^2 \Kzero_i(k,p)
= \Slash{k}\rho^0(k)P^T_{ij}(k)(\nb(k^0) + \nf(k^0))\\
&\times\cp \Cf\left(\gamma^{j}\varPsi(p)+\cp \int \frac{d^4 k'}{(2\pi)^4} \frac{k^k\gamma^j+\gamma^kk'{}^{j}}{k\cdot k'}  \Kzero_k(k',p)\right).
\end{split}
\end{align}
To solve this self-consistent equation, we assume that the off-diagonal propagator at zeroth order has the following structure: 
\begin{align}
\begin{split}  
&\Kzero_i(k,p) \\
&= \frac{\cp\Cf A\Slash{k}}{\delta m^2}\rho^0(k)P^T_{ij}(k)(\nb(k^0) + \nf(k^0)) \gamma^{j}\varPsi(p),
\end{split}
\end{align}
where $A$ is a constant.
By utilizing this assumption, the following expression is obtained~\cite{ultrasoft-QED}:
\begin{align} 
\begin{split}
&\int \frac{d^4 k'}{(2\pi)^4} \frac{k^k\gamma^j+\gamma^kk'{}^{j}}{k\cdot k'}  \Kzero_k(k',p) \\
&\simeq -A\Cf\gamma^j\frac{\cp T^2}{8\delta m^2}\varPsi(p),
\end{split}
\end{align}
where we have used that $k$ is on-shell in Eq.~(\ref{eq:app-ultrasoft-1}) and neglected the term that is proportional to $\Slash{k}$, which is justified since its contribution to Eq.~(\ref{eq:app-ultrasoft-1}) is negligible due to the presence of another $\Slash{k}$.
By using this expression, Eq.~(\ref{eq:app-ultrasoft-1}) can be solved and the result is
\begin{align} 
\label{eq:A-result}
 A&= \frac{1}{1+\lambda}.
\end{align}

Collecting the first order terms in Eq.~(\ref{eq:EOM-result-linear-response-color}) yields the following equation:
\begin{align}
\begin{split} 
& 2k\cdot \tilde{p}\Kzero_i(k,p)+\delta m^2\Kone_i(k,p)\\
&= \cp^2 \Cf\Slash{k}\rho^0(k)P^T_{ij}(k)(\nb(k^0) + \nf(k^0))\\
&\times\int \frac{d^4 k'}{(2\pi)^4} \frac{k^k\gamma^j+\gamma^kk'{}^{j}}{k\cdot k'}  \Kone_k(k',p).
\end{split}
\end{align}
Instead of solving this equation, we multiply this equation by $\gamma^i$ and integrate over $k$.
Then, the equation becomes
\begin{align}
\begin{split} 
&\int \frac{d^4 k}{(2\pi)^4} 2k\cdot \tilde{p}\gamma^i\Kzero_i(k,p)+\delta m^2\int \frac{d^4 k}{(2\pi)^4}\gamma^i\Kone_i(k,p)\\
&= \cp^2 \Cf\int \frac{d^4 k}{(2\pi)^4}\Slash{k}\rho^0(k)\gamma^iP^T_{ij}(k)(\nb(k^0) + \nf(k^0))\\
&\times\int \frac{d^4 k'}{(2\pi)^4} \frac{k^k\gamma^j+\gamma^kk'{}^{j}}{k\cdot k'}  \Kone_k(k',p).
\end{split}
\end{align}
By utilizing the properties~\cite{ultrasoft-QED}
\begin{align}
\begin{split}
&\int \frac{d^4 k}{(2\pi)^4} k\cdot \tilde{p}\gamma^i\Kzero_i(k,p) 
=\frac{A\cp\Cf\pi^2T^4}{8\delta m^2 }Y(\tilde{p})\varPsi(p),
\end{split}
\end{align}
 and 
\begin{align}
\begin{split}
&\int \frac{d^4 k}{(2\pi)^4}\Slash{k}\rho^0(k)\gamma^iP^T_{ij}(k)(\nb(k^0) + \nf(k^0))\\
&~~~\times\int \frac{d^4 k'}{(2\pi)^4} \frac{k^k\gamma^j+\gamma^kk'{}^{j}}{k\cdot k'}  \Kone_k(k',p) \\
&\simeq-\frac{T^2}{8} \int \frac{d^4 k'}{(2\pi)^4} \gamma^k\Kone_k(k',p),
\end{split}
\end{align}
we get
\begin{align}
\begin{split}
\int \frac{d^4 k}{(2\pi)^4}\gamma^i\Kone_i(k,p)
&\simeq -\frac{\cp \Cf A^2\pi^2T^4}{4(\delta m^2)^2} Y(\tilde{p})\varPsi(p) .
\end{split}
\end{align}

Thus, we can calculate the induced fermionic source using Eq.~(\ref{eq:etaind-K}):
\begin{align}
\label{eq:app-ultrasoft-etaind-result}
\etaind(p)&\simeq -\frac{Y(\tilde{p})\varPsi(p)}{Z}.
\end{align}
Here we have used the fact that the integral of $\Kzero_i(k,p)$ vanishes because it is odd function of $k$.
By remembering Eq.~(\ref{eq:selfenergy-current}), we can obtain the quark self-energy with a ultrasoft momentum, which is given in Eq.~(\ref{eq:fermion-mode-result}), from the induced fermionic source given in Eq.~(\ref{eq:app-ultrasoft-etaind-result}).

We note that the same result can be obtained from the resummed perturbation theory~\cite{lebedev, ultrasoft-QED} because that scheme is equivalent to the GBE.

%%%%%%%%%%%%%%%%%%%%%%%%%%%%%%%%%%%%%%%%%%%%%%%%%%%%%%%%%%%%%%%%%%%

\end{document}